\documentclass[12pt, runningheads]{llncs}
\usepackage{graphicx}

\begin{document}
\title{How Hard Is Squash? - Towards Information Theoretic Analysis of Motor Behavior in Squash}
%
%

%

\titlerunning{How Hard Is Squash?}

\author{ Kavya Anand\and
Pramit Saha}
\authorrunning{Anand et al.} 

%
\maketitle              
\begin{abstract}
Fitts' law has been widely employed as a research method for analyzing tasks within the domain of Human-Computer Interaction (HCI). However, its application to non-computer tasks has remained limited. This study aims to extend the application of Fitts' law to the realm of sports, specifically focusing on squash. Squash is a high-intensity sport that requires quick movements and precise shots. Our research investigates the effectiveness of utilizing Fitts' law to evaluate the task difficulty and effort level associated with executing and responding to various squash shots. By understanding the effort/information rate required for each shot, we can determine which shots are more effective in making the opponent work harder. Additionally, this knowledge can be valuable for coaches in designing training programs. However, since Fitts' law was primarily developed for human-computer interaction, we adapted it to fit the squash scenario. This paper provides an overview of Fitts' law and its relevance to sports, elucidates the motivation driving this investigation, outlines the methodology employed to explore this novel avenue, and presents the obtained results, concluding with key insights. We conducted experiments with different shots and players, collecting data on shot speed, player movement time, and distance traveled. Using this data, we formulated a modified version of Fitts' law specifically for squash. The results provide insights into the difficulty and effectiveness of various shots, offering valuable information for both players and coaches in the sport of squash.
\end{abstract}

\section{Introduction}

Squash \cite{jones2018review} is a fast-paced and physically demanding sport that requires exceptional agility, quick reflexes, and precise shot-making abilities. Played on a small, enclosed court with four walls and a ceiling, squash provides a unique challenge for players as they navigate the confined space and strategically outmaneuver their opponents. The combination of intense movements, such as running, lunging, and jumping, makes squash a thrilling and captivating sport to play and watch.

In a game of squash, players employ various shots to gain control of the rally and outwit their opponents. Each shot carries its own purpose and technique, adding an element of strategy and unpredictability to the game. The drive, one of the most common shots, involves hitting the ball straight at a medium to low height along the wall, aiming to maintain ball control and restrict the opponent's reach. The boast, on the other hand, entails hitting the ball off the side wall at an angle, causing it to bounce towards the front wall and away from the opponent, often surprising them and altering the rally's direction. The drop shot requires delicate touch and precision, landing the ball softly near the front wall, just out of the opponent's reach. The lob, a defensive shot, involves hitting the ball high and deep into the back of the court, buying the player time to recover and prepare for the next shot. 

Understanding the dynamics and complexities of squash shots is essential for players to excel in the sport. It not only requires technical proficiency but also demands a deep comprehension of movement efficiency and decision-making. In this regard, the application of Fitts' law in squash can provide valuable insights into the biomechanics and cognitive processes involved in shot selection and execution.

Fitts' law, originally formulated for human-computer interaction and small tasks on digital devices, offers a framework to quantify the relationship between movement time, target difficulty, and motor control. By adapting Fitts' law to the context of squash, we can measure the movement time required to reach and return various shots and calculate an index of difficulty for each shot. This allows us to determine the effort/information rate associated with different shot movements. Exploring the application of Fitts' law in squash has significant implications for the field of sports, particularly in understanding the efficiency and effectiveness of different shots. It provides valuable insights into optimizing player training and coaching methodologies by identifying shots that demand additional focus and practice.

However, it is important to note that the original formulation of Fitts' law may require adjustments to accommodate the unique characteristics of squash shots and movements. Therefore, we conducted a series of experiments, analyzing players of varying skill levels retrieving different shots. Through data analysis and statistical modeling, we formulated and validated a modified Fitts' law equation specific to squash, capturing the relationship between movement time, index of difficulty, and information processing rate.

In this paper, we present the findings of our study on the application of Fitts' law in squash. We delve into the background of Fitts' law, discussing its original formulation and close variants that address certain limitations. We then outline the problem formulation, explaining how we adapted Fitts' law to measure the difficulty and effort of picking up squash shots. The methodology section details our experimental setup, data collection process, and the calculations used to determine the index of difficulty and movement time for each shot. Finally, we present the results and discuss their implications for players, coaches, and the overall understanding of squash as a sport.

By leveraging the principles of Fitts' law in the context of squash, this research aims to enhance our understanding of shot dynamics, performance optimization, and strategic decision-making. Ultimately, it contributes to the ongoing exploration of sports science and human performance in the realm of squash, providing a foundation for further research and practical applications in training and competition. 

To the best of our knowledge, this is the first work to systematically investigate the application of Fitts' law in the context of squash shots. By adapting this well-established framework to the unique demands of squash, we have made significant strides in understanding the biomechanics, cognitive processes, and efficiency of the movement required to retrieve different shots. This study represents a considerable effort in bridging the gap between sports science and squash, offering valuable insights that can inform training methodologies, enhance player performance, and contribute to the overall advancement of the sport.

\begin{figure*}

    \includegraphics[width=1\columnwidth]{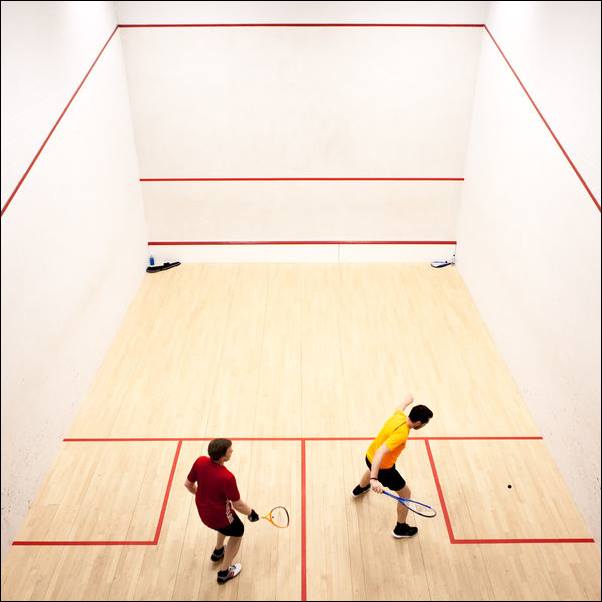}
    \caption{Dynamic gameplay in a Squash Court: Capturing the intensity of competitive exchange between two skilled players.}
    \label{fig:1}
\end{figure*}

\section{Background}
\subsection{Fitts' experiment}
Fitts' original experiment \cite{fitts1954information}, conducted by Paul M. Fitts, played a pivotal role in formulating the fundamental equation that expresses the relationship between throughput, the index of difficulty (ID), and movement time (MT). The experiment involved a simple tapping task, wherein two targets of width W were placed at a distance A from each other. By measuring the time taken to tap each target alternately, Fitts calculated the index of difficulty using the equation: $ID = \log _2 (2A/W)$. 

In Fitts' experiment, the movement time ranged from 180ms to 731ms. The index of performance for various experimental setups was calculated by dividing the calculated ID by the movement time. Fitts observed that the mean index of performance was 10.10 bits/s, with a standard deviation of 1.33 bits/s, which he assumed to be the average information processing rate of the human motor control system.

In the original Fitts' law formulation, the focus was primarily on predicting movement time based on the difficulty of the task represented by the index of difficulty (ID), which is calculated using the target width (W) and distance (A) between targets. The term throughput was defined to incorporate the concept of information processing rate or task performance by taking into account the trade-off between speed and accuracy in motor tasks. It provides a more comprehensive understanding of the relationship between movement time, index of difficulty, and task performance. The throughput (TP) in the context of Fitts' law refers to the rate at which information is processed and tasks are completed. It is calculated by dividing the index of difficulty (ID) by the movement time (MT), resulting in the formula: $TP = ID/MT$.

The throughput model recognizes that maximizing speed alone may compromise accuracy, while maximizing accuracy alone may result in slower task completion. By considering the throughput, researchers can evaluate the efficiency of a task performance, taking into account both speed and accuracy. It provides insights into the information processing capabilities of individuals and can be used to compare performance across different tasks or conditions. A higher throughput value indicates a more efficient performance, as it reflects a greater amount of information processed per unit of time. The throughput model has been widely used in the field of human-computer interaction, ergonomics, and user interface design. It helps in understanding the impact of task characteristics, such as target size and distance, on the overall performance of motor tasks. By considering throughput, researchers and designers can optimize task conditions to achieve the desired balance between speed and accuracy.

\begin{figure*}
    \includegraphics[width=1\columnwidth]{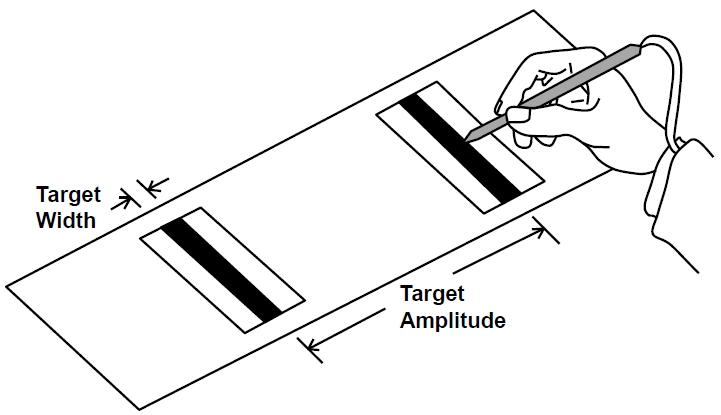}
    \caption{Fitts' Original Experiment}
    \label{fig:2}
\end{figure*}

\subsection{Close Variants of Fitts' law}

While Fitts' law demonstrated a high correlation between movement time and index of difficulty, there were some issues encountered. As the values for the index of difficulty decreased, the movement time deviated from the regression line, indicating non-compliance with Fitts' formulation. To address this, variants of Fitts' law were proposed. 

\textbf{Mackenzie's Variant: \cite{mackenzie1992fitts}} One variant, introduced by Scott Mackenzie, adjusted the movement time formula to $MT = a + \log_2 (2A/W + 1)$, ensuring that as the value of $2A/W$ approached 0, the index of difficulty remained non-negative. This adjustment resulted in a more practical formulation and improved the alignment of the graph for lower index of difficulty values with the regression line. 

\textbf{Welford formulation: \cite{mackenzie1993fitts}} Another variant, known as the Welford formulation, introduced the equation $MT = a + b_1 \log 2(A) + b_2 \log_2 (1/W)$, which can be generalized to $ID = b \log_2 (a/W_k)$.

\textbf{Zhai-Accot's Steering law: \cite{accot1997beyond} .\cite{accot1999performance}}
Additionally, Zhai and Accot developed the steering law as an extension of Fitts' law to account for trajectory tasks, as Fitts' law is primarily applicable to pointing tasks. The steering law considers both the targets and the path between them, with the path width influencing the task difficulty. This law modified the index of difficulty calculation by using the ratio of $ A/W$, representing the path width, in the equation $MT = a + b (A/W)$.


\textbf{Multiple-Target Fitts' Law: \cite{balakrishnan2004beating}} The multiple-target Fitts' law extends the original formulation to tasks involving multiple targets. It accounts for the complexity and characteristics of tasks that require multiple sequential movements. This variant considers the total distance traveled and the time taken to complete all the movements in the task.

\textbf{Extended Fitts' Law: \cite{mackenzie1992extending}} This variant incorporates additional factors such as target distance, target amplitude, and target width to provide a more comprehensive model for predicting movement time. It takes into account the movement trajectory and the complexity of the task, resulting in improved accuracy in predicting performance.

\textbf{Multi-directional Fitts' Law: \cite{mackenzie2012fittstilt}} The original Fitts' law assumes unidirectional movements, but in many real-world tasks, movements occur in multiple directions. This variant extends Fitts' law to include movements in different directions, considering the angular aspect of the task. It provides a more accurate prediction of movement time for tasks involving movements in various directions.

\textbf{Discrete Fitts' Law: \cite{jagacinski1980fitts} .\cite{guiard1997fitts}} Discrete Fitts' law is a variant that applies Fitts' law to discrete pointing tasks, where the targets are presented one at a time. It is particularly relevant in situations where sequential targeting is required, such as menu selection or scrolling. This variant considers the time required to transition between targets, improving the applicability of Fitts' law to discrete tasks.

\textbf{Compound Fitts' Law: \cite{milner1992model}} Compound Fitts' law extends the original formulation to accommodate tasks with multiple subtasks or sequential movements. It considers the time required to complete each subtask and combines them to estimate the overall movement time. This variant is useful in analyzing complex tasks that involve a series of discrete movements.

\textbf{Model-based Fitts' Law: \cite{crossman1983feedback}} Model-based variants of Fitts' law incorporate additional parameters based on specific models of human movement. These models may take into account factors such as acceleration, deceleration, and target acquisition dynamics to improve the accuracy of predicting movement time. Model-based approaches enhance the applicability of Fitts' law to a wide range of tasks and movement scenarios.

\textbf{Expanding Targets Fitts' Law: \cite{mcguffin2005fitts}} This variant extends Fitts' law to accommodate targets that dynamically change in size during the task. Instead of a fixed target width, the target width continuously expands or contracts based on certain criteria. This variant is particularly relevant in interactive user interfaces where targets may dynamically resize based on user input or system feedback.

\textbf{Time-Constrained Fitts' Law:} \cite{zelaznik1988role} This variant incorporates time constraints into the Fitts' law formulation. It aims to predict the achievable throughput or performance under specific time limitations. By considering the movement time and index of difficulty along with a predefined time constraint, this variant provides insights into the optimal performance that can be achieved within a given time frame.

\textbf{Bivariate Fitts' Law: \cite{accot2003refining}} Bivariate Fitts' law extends the original formulation to include additional parameters that account for the task's spatial and temporal aspects. It considers factors such as movement direction, movement speed, and the correlation between spatial and temporal characteristics of the task. This variant provides a more comprehensive model that captures the complex relationship between movement parameters and performance.

\textbf{Cognitive Fitts' Law: \cite{sleimen2013age} } Cognitive Fitts' law expands the original formulation by incorporating cognitive factors into the model. It takes into account cognitive load, attentional demands, and decision-making processes during the task. By considering the cognitive aspects of motor control, this variant provides insights into the influence of cognitive factors on movement performance and efficiency.

\textbf{Fitts' Law in 3D Space: \cite{cha2013extended,zeng2012fitts}} While the original Fitts' law primarily applies to 2D pointing tasks, variants have been proposed to extend it to three-dimensional (3D) space. These variants consider additional dimensions, such as depth or height, along with horizontal and vertical movements. They account for the spatial characteristics of tasks that involve reaching or pointing in 3D environments, such as virtual reality or robotic applications.

These variants and extensions of Fitts' law provide researchers with a range of tools to account for different task characteristics and improve the applicability of the model in various domains. 

\section{Problem formulation, solution overview, and assumptions}

In this study, we aimed to apply Fitts' law to the specific scenario of picking up various shots in our game. However, since the conventional formulation of Fitts' law is primarily designed for pointing tasks, we recognized the need for a modified version tailored to our unique situation. To achieve this, we conducted a comprehensive analysis of the factors influencing the difficulty of picking up shots and performed a pilot study in the squash court as a sanity test to verify its applicability.

Two key factors emerged as crucial determinants: the speed of the ball being played and the distance that players need to cover to reach the desired shot location. These factors directly impact the complexity and challenge involved in successfully picking up shots.
The speed of the ball influences the player's reaction time and the level of precision required to intercept the shot effectively. Faster-moving balls demand quicker reflexes and more accurate positioning, increasing the overall difficulty of shot selection.
Additionally, the distance that players need to travel to reach the desired shot location affects the time available for movement and the positioning required to retrieve the ball effectively. Longer distances can lead to more challenging shot pickups, as they require swift and efficient movement across the court.

By incorporating these factors into our index of difficulty formulation, we can quantitatively assess the level of challenge associated with different shot selections. Our modified version of Fitts' law takes into account the speed of the ball (v) and the distance to the shot location (D), resulting in the equation:

\begin{equation}
ID = \log _2 (vD)
\end{equation}

Here, $v$ represents the speed of the ball, and $D$ corresponds to the distance the player must cover to successfully pick up the shot. This equation allows us to calculate an index of difficulty (ID) specific to our gameplay scenario for each shot, providing a measure of the relative difficulty involved in successfully picking up different shots in our squash game.

To further assess player performance, we integrated the calculated ID value into the Fitts' law equation, giving us the index of performance (IP):
\begin{equation}
IP = ID / MT = \log _2 (vD) / MT
\end{equation}
In this equation, 'MT' denotes the time it takes for the player to reach the shot and successfully pick it up. By applying this modified formulation, we aim to gain insights into the relationship between the difficulty of shot selection, player movement time, and overall performance.

Apart from the speed of the ball and the distance that players need to travel, there are several other factors that can influence the difficulty of picking up squash shots. In this experiment, we ensure the controlled manipulation of these factors to maintain consistency in the difficulty level, minimizing their impact on the results. However, if the factors vary significantly, one should consider these factors alongside the speed of the ball and distance to the shot location to get a more comprehensive and accurate assessment of the difficulty involved in picking up squash shots. These factors include:

\begin{enumerate}

\item \textbf{Shot Angle:} The angle at which the ball is played can affect the difficulty of shot selection. Shots played at acute angles or extreme angles away from the player's position may require more precise movement and positioning, increasing the overall difficulty.

\item \textbf{Shot Placement:} The location on the court where the ball is played can impact the complexity of shot retrieval. Shots placed near the corners or close to the wall may require players to cover a larger distance or adopt challenging body positions, adding to the difficulty level.

\item  \textbf{Shot Velocity:} In addition to the speed of the ball, the velocity at which the ball is struck by the opponent can influence the difficulty. Higher ball velocities demand quicker reactions and increased agility to intercept the shot effectively.

\item \textbf{Shot Spin:} The spin or rotation of the ball can introduce unpredictability and variation in shot trajectories. Dealing with spin requires players to adjust their positioning, footwork, and racket angle, thereby affecting the difficulty of shot pickup.

\item \textbf{Player Positioning:} The initial position of the player on the court relative to the shot location plays a crucial role in shot retrieval difficulty. If the player is out of position or too far from the desired shot location, it will require more effort and time to reach and retrieve the ball.

\item \textbf{Player Fitness and Stamina:} The physical fitness and stamina of the player can influence their ability to quickly cover distances and maintain performance over extended periods. Fatigue can increase the difficulty of shot pickups, especially in demanding gameplay situations.

\item \textbf{Environmental Factors:} External factors such as lighting conditions, court surface, temperature, and humidity can also impact shot retrieval difficulty. Poor visibility, slippery surfaces, or adverse weather conditions can pose additional challenges for players.

\end{enumerate}

\section{Methodology}

\subsection{Participants}
The experiment involved three subjects who were proficient in the sport of squash, each representing a different skill level. The participants were selected based on their experience and expertise in playing squash to ensure a range of skill levels and to capture a comprehensive understanding of the factors influencing shot difficulty.

\subsection{Experimental Design}
The experiment focused on four different types of squash shots, aiming to evaluate the difficulty level of picking up these shots. Each participant performed three trials for each shot, resulting in a total of 12 trials per participant. The shots were performed by a designated player positioned at the center of the court, while the subjects, also positioned at the center, were responsible for picking up the shots. The experiment was carried out sequentially for each participant, beginning with the young participant with expert skill level, followed by the old participant with expert skill level, and concluding with the young participant with a moderate skill level.

\subsection{Court Configuration}
The experiment took place in a standard squash court with well-defined dimensions conforming to the official standards of the sport. The center court area, where the shots were played and picked up, is located 4.2 meters from the back wall and 5.55 meters from the front wall. It is situated between the two side walls, with a distance of 3.2 meters from each side. This central area forms a T-shaped region, as depicted in Figure 3.

\subsection{Experimental Instruments}
The squash racket used to play all the shots was the Tecnifibre Carboflex 125 X-Speed. The video recording device used was an iPhone 13 and the video editing software (whose use is highlighted later) used was EaseUS. To measure the distances, the in-built measurement software on the iPhone 13 was used.

\subsection{Data Collection and Analysis}
To capture relevant data for analysis, a video recording of the experiment was taken by a designated observer positioned either at the back or to the side of the court. This recording allowed for a detailed post-experiment analysis.

After picking up each shot, the player maintained a stationary position with the racket in the same position where it made contact with the ball. The distance between the player's starting position and the pickup location was measured and recorded as the distance traveled (D) by the player to reach the shot.

The video analysis involved determining the point of contact between the ball and the front wall. The distance between the player who hit the shot and the point of contact on the front wall was measured to evaluate the distance travelled by the ball to eventually calculate its velocity.

Furthermore, the video recording was slowed down by a factor of 10 to enable precise measurement of the time taken by the ball to reach the front wall. This, along with the calculation of distance travelled to the front wall mentioned above, allowed for the calculation of the ball's speed (v). It is important to note that, due to the short duration of the time period, the ball's average speed was assumed to be equal to its instantaneous speed.

\subsection{Data Analysis and Calculations}
The collected data, including the distance traveled by the player (D) and the speed of the ball (v), were used to calculate the index of difficulty (ID) for each shot as expressed in Equation 1.

Subsequently, the obtained value of ID was utilized in the original Fitts' law equation to determine the index of performance (IP) for each shot. The index of performance was calculated using the Equation 2.

The resulting data and performance metrics provided valuable insights into the relationship between shot difficulty, player movement, and ball characteristics, contributing to a comprehensive analysis of the factors influencing the difficulty of picking up squash shots.

\textbf{Note:} Ethical guidelines and informed consent protocols were followed throughout the experiment to ensure the well-being and rights of the participants.
\\ 
\begin{figure*}
    \includegraphics[width=1\columnwidth]{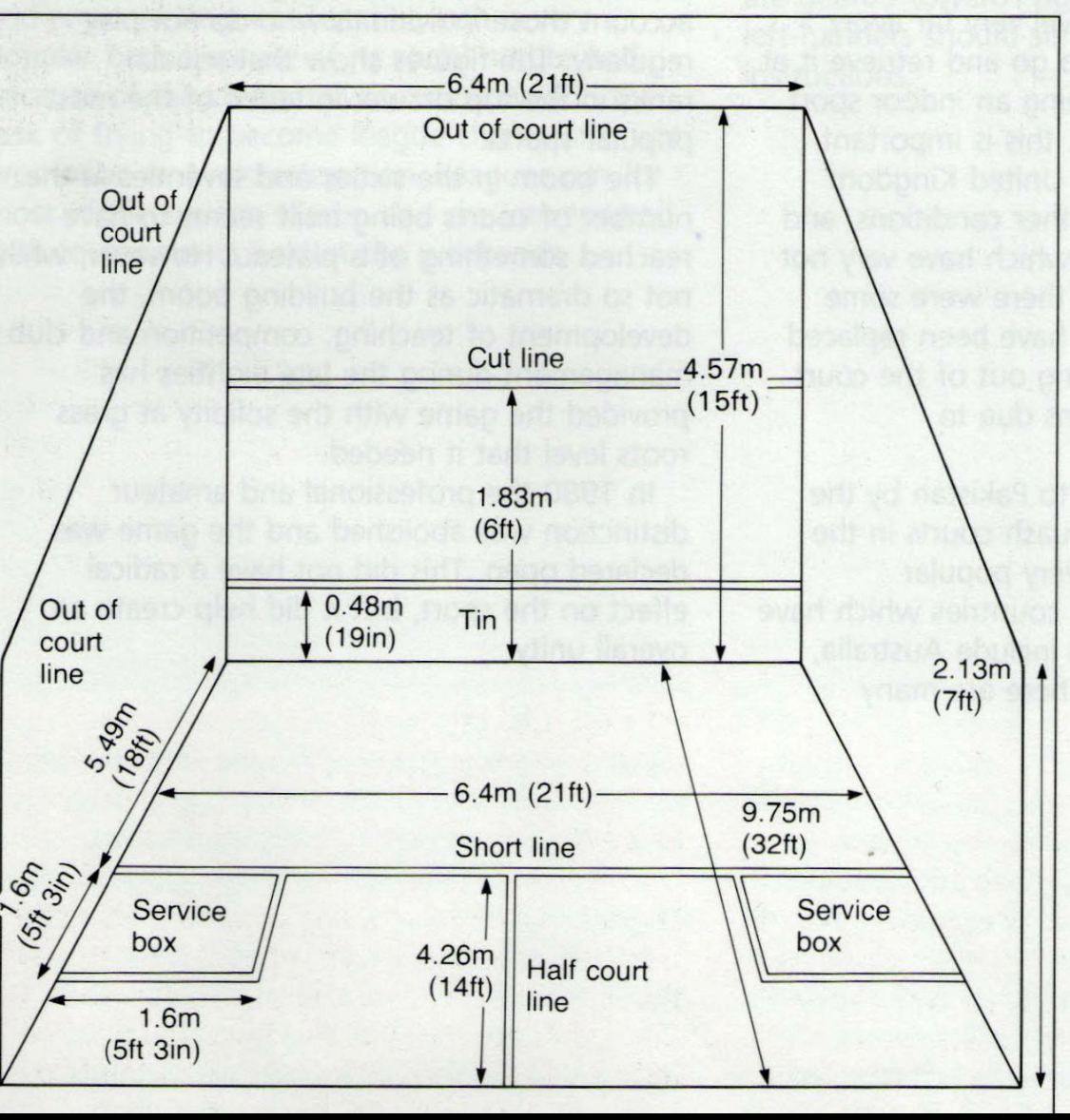}
    \caption{Task Space: Dimensions of a Squash Court}
    \label{fig:3}
\end{figure*}
\\ 
\section{Results}

\begin{table}[!ht]
    \centering
    \caption{Data from the Squash Experiment on Fitts' law - DB refers to the distance travelled by ball, T refers to the time taken by the ball to cover that distance, v refers to the speed of the ball, DP refers to the distance travelled by the player, ID refers to the index of difficulty, MT refers to the movement time of the player and IR refers to the information rate}
    \begin{tabular}{|l|l|l|l|l|l|l|l|l|l|}
    \hline
        \textbf{Person} & \textbf{Shot} & \textbf{Trial} &   \textbf{DB(cm)} & \textbf{T (s)} & \textbf{V(m/s)} & \textbf{DP (cm)} & \textbf{ID (bits)} & \textbf{MT(s)} & \textbf{IR (bits/s)}  \\ \hline \hline
        1 & Drive & 1 & 586 & 0.197 & 29.75 & 374 & 6.8 & 1.22 & 5.57  \\ \hline
        1 & Drive & 2 & 587 & 0.204 & 28.77 & 386 & 6.8 & 1.21 & 5.62  \\ \hline
        1 & Drive & 3 & 580 & 0.198 & 29.29 & 454 & 7.01 & 1.23 & 5.7  \\ \hline
        1 & Drop & 1 & 615 & 0.395 & 15.57 & 355 & 5.79 & 1.06 & 5.46  \\ \hline
        1 & Drop & 2 & 614 & 0.3775 & 16.26 & 352 & 5.84 & 1.09 & 5.36  \\ \hline
        1 & Drop & 3 & 605 & 0.395 & 15.32 & 364 & 5.8 & 1.04 & 5.58  \\ \hline
        1 & Lob & 1 & 686 & 0.3125 & 21.95 & 380 & 6.38 & 1.89 & 3.38  \\ \hline
        1 & Lob & 2 & 665 & 0.3625 & 18.34 & 402 & 6.2 & 1.63 & 3.8  \\ \hline
        1 & Lob & 3 & 670 & 0.3275 & 20.46 & 361 & 6.21 & 1.78 & 3.49  \\ \hline
        1 & Boast & 1 & 971 & 0.792 & 12.26 & 491 & 5.91 & 1.013 & 5.83  \\ \hline
        1 & Boast & 2 & 980 & 0.732 & 13.39 & 360 & 5.59 & 1.208 & 4.63  \\ \hline
        1 & Boast & 3 & 961 & 0.715 & 13.44 & 350 & 5.56 & 1.27 & 4.38  \\ \hline \hline
        2 & Drive & 1 & 598 & 0.258 & 23.18 & 386 & 6.48 & 1.361 & 4.76  \\ \hline
        2 & Drive & 2 & 567 & 0.21 & 27 & 388 & 6.71 & 1.256 & 5.34  \\ \hline
        2 & Drive & 3 & 593 & 0.204 & 29.07 & 316 & 6.52 & 1.321 & 4.94  \\ \hline
        2 & Drop & 1 & 636 & 0.446 & 14.26 & 260 & 5.21 & 1.12 & 4.65  \\ \hline
        2 & Drop & 2 & 679 & 0.47 & 14.45 & 330 & 5.58 & 1.08 & 5.17  \\ \hline
        2 & Drop & 3 & 587 & 0.413 & 14.21 & 285 & 5.34 & 1.11 & 4.81 \\ \hline
        2 & Lob & 1 & 686 & 0.283 & 24.24 & 308 & 6.22 & 1.45 & 4.29  \\ \hline
        2 & Lob & 2 & 720 & 0.276 & 26.09 & 362 & 6.56 & 1.89 & 3.47  \\ \hline
        2 & Lob & 3 & 710 & 0.335 & 21.19 & 329 & 6.12 & 1.46 & 4.19  \\ \hline
        2 & Boast & 1 & 982 & 1.02 & 9.63 & 469 & 5.5 & 1.35 & 4.07  \\ \hline
        2 & Boast & 2 & 999 & 0.964 & 10.36 & 425 & 5.46 & 1.63 & 3.35  \\ \hline
        2 & Boast & 3 & 991 & 0.94 & 10.54 & 440 & 5.54 & 1.42 & 3.9  \\ \hline \hline
        3 & Drive & 1 & 625 & 0.201 & 31.09 & 332 & 6.69 & 1.322 & 5.06  \\ \hline
        3 & Drive & 2 & 596 & 0.196 & 30.41 & 351 & 6.74 & 1.34 & 5.03  \\ \hline
        3 & Drive & 3 & 610 & 0.24 & 25.42 & 367 & 6.54 & 1.27 & 5.15  \\ \hline
        3 & Drop & 1 & 711 & 0.452 & 15.73 & 373 & 5.87 & 1.09 & 5.39  \\ \hline
        3 & Drop & 2 & 690 & 0.411 & 16.79 & 350 & 5.88 & 1.081 & 5.44  \\ \hline
        3 & Drop & 3 & 662 & 0.43 & 15.4 & 360 & 5.79 & 1.113 & 5.2  \\ \hline
        3 & Lob & 1 & 600 & 0.291 & 20.62 & 360 & 6.21 & 1.78 & 3.49  \\ \hline
        3 & Lob & 2 & 640 & 0.334 & 19.16 & 371 & 6.15 & 1.64 & 3.75  \\ \hline
        3 & Lob & 3 & 701 & 0.29 & 24.17 & 388 & 6.55 & 1.52 & 4.31  \\ \hline
        3 & Boast & 1 & 937 & 0.999 & 9.38 & 483 & 5.5 & 1.13 & 4.87  \\ \hline
        3 & Boast & 2 & 962 & 0.91 & 10.57 & 460 & 5.6 & 1.56 & 3.59  \\ \hline
        3 & Boast & 3 & 969 & 1.08 & 8.97 & 490 & 5.46 & 1.44 & 3.79 \\ \hline \hline
    \end{tabular}
\end{table}

In this section, we report the experimental results depicting the relationship between index of difficulty (ID) and movement time (MT) along with the corresponding index of performance/information rate for each person and each shot. 

For Person 1, the average ID for drives was 6.87 bits (SD = 0.1), while the mean ID for drops was 5.81 bits (SD = 0.02). This indicates that, on average, the drop shot is easier to retrieve. Similarly, the mean ID for lobs was 6.26 bits (SD = 0.08), and for boasts, it was 5.69 bits (SD = 0.16), suggesting that boasts were the easiest shot for Person 1.

For Person 2, the average ID for drives was 6.57 bits (SD = 0.1), while the average ID for drops was 5.38 bits (SD = 0.15). The average ID for lobs was 6.3 bits (SD = 0.2), and for boasts, it was 5.5 bits (SD = 0.03).

For Person 3, the average ID for drives was 6.66 bits (SD = 0.08), while the mean ID for drops was 5.85 bits (SD = 0.04). The mean ID for lobs was 6.3 bits (SD = 0.2), and for boasts, it was 5.52 bits (SD = 0.06).

When considering all the players, the mean ID for drives was 6.7 bits (SD = 0.16), for drops it was 5.68 bits (SD = 0.23), for lobs it was 6.29 bits (SD = 0.16), and for boasts it was 5.57 bits (SD = 0.13).

The correlation between MT and ID is illustrated in the graph below. The positive correlation is evident from the upward sloping trend line, indicating that as ID increases, so does MT. The data points are scattered around the trend line but remain relatively close to it, confirming the relationship between MT and ID. The equation of the trend line is given as MT = 0.456 ID - 1.3. 

\begin{figure*}
    \includegraphics[width=1\columnwidth]{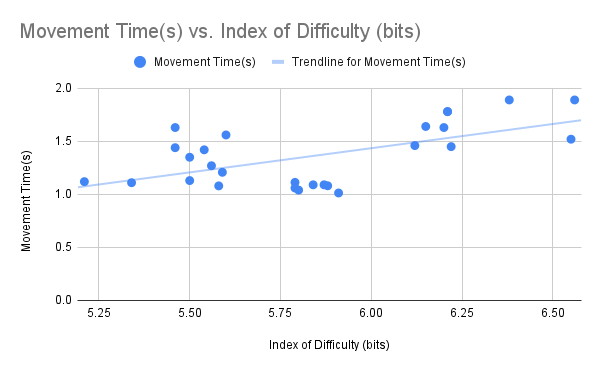}
    \caption{Overall Relationship Movement Time (s) vs Index of Difficulty (bits)}
    \label{fig:4}
\end{figure*}

\begin{figure*}
    \includegraphics[width=1\columnwidth]{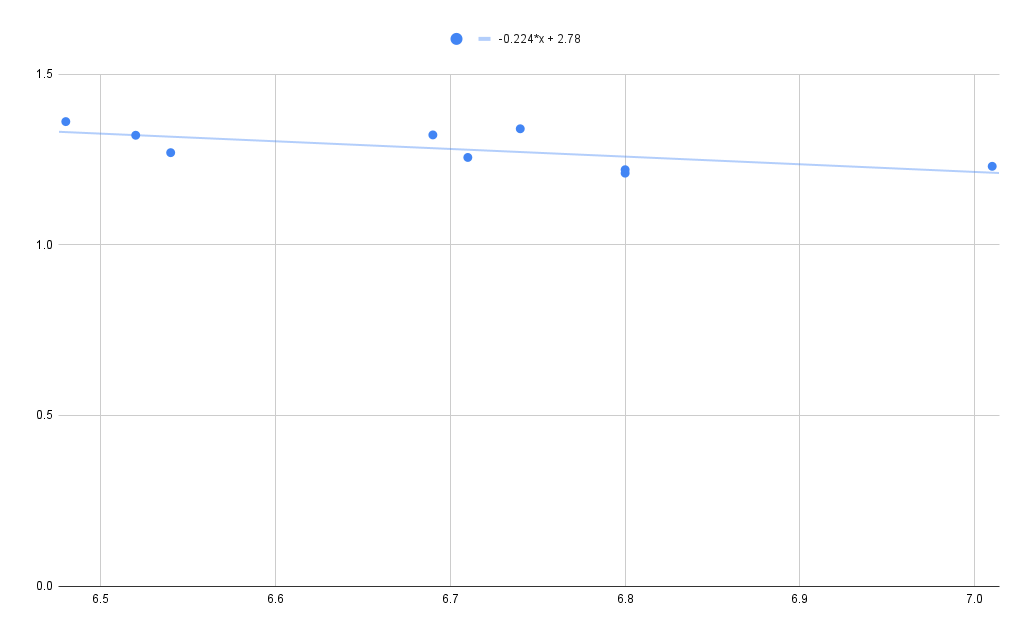}
    \caption{Movement Time (s) vs Index of Difficulty (bits) for drives}
    \label{fig:5}
\end{figure*}

\begin{figure*}
    \includegraphics[width=1\columnwidth]{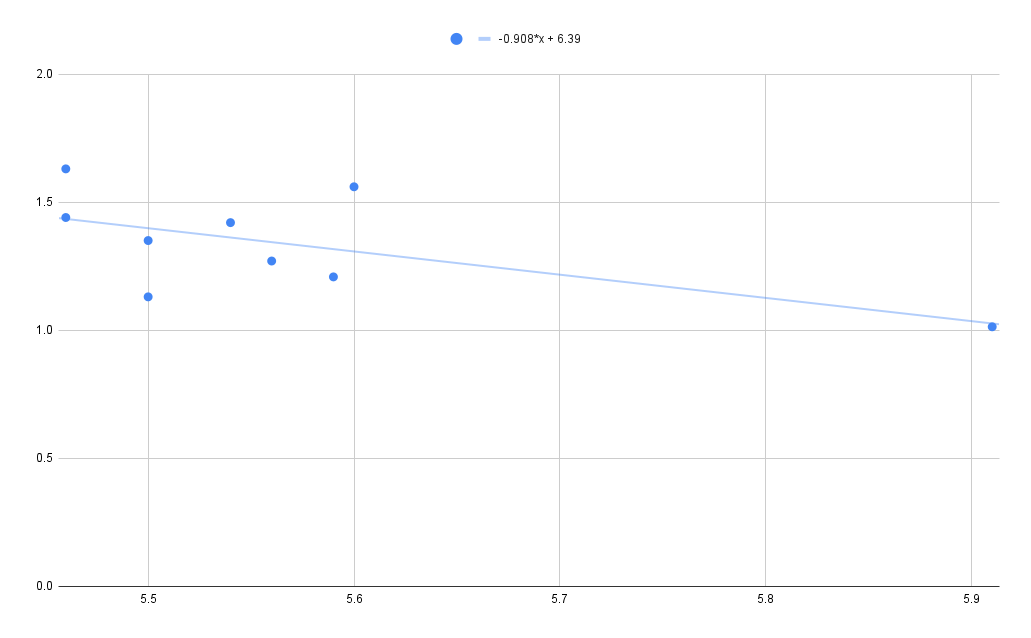}
    \caption{Movement Time (s) vs Index of Difficulty (bits) for boasts}
    \label{fig:6}
\end{figure*}

\begin{figure*}
    \includegraphics[width=1\columnwidth]{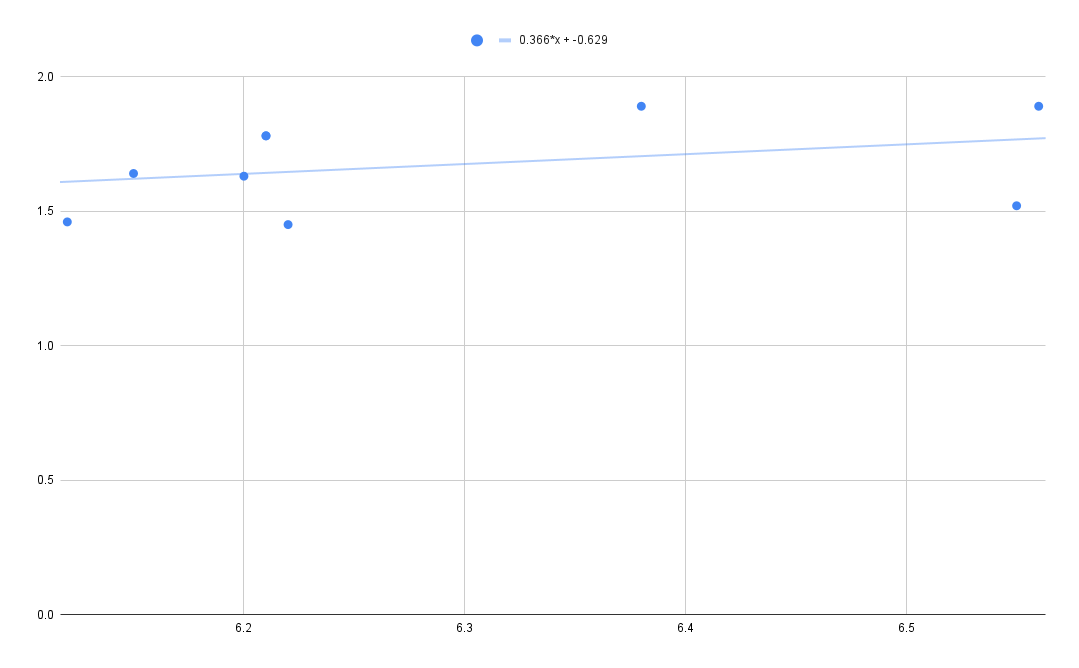}
    \caption{Movement Time (s) vs Index of Difficulty (bits) for lobs}
    \label{fig:7}
\end{figure*}

\begin{figure*}
    \includegraphics[width=1\columnwidth]{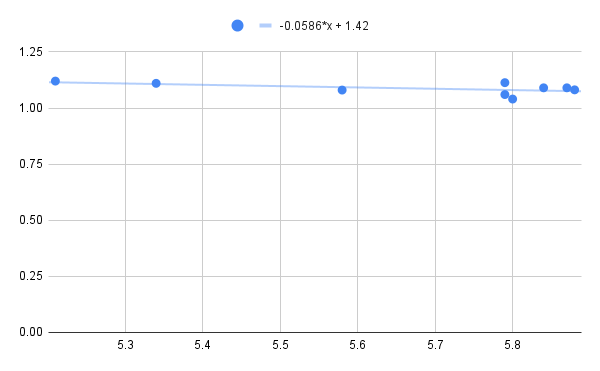}
    \caption{Movement Time (s) vs Index of Difficulty (bits) for drops}
    \label{fig:8}
\end{figure*}

\section{Discussions}
\subsection{Implications of Index of Difficulty}
The findings indicate that the Index of Difficulty (ID) is a reasonable measure that effectively captures the difficulty of squash shots. The low standard deviation in the average ID suggests that the formulation of ID adequately considers the essential factors influencing shot difficulty, specifically the ball speed and its landing position. The ID appears to be primarily influenced by these factors, which aligns with the intuitive understanding that better shots are designed to challenge opponents by increasing the time required to reach them.
\subsection{Overall relationship between movement time and index of difficulty}

The increasing index of difficulty (ID) indicates an improvement in the quality of shots played which makes it harder to pick up, thus resulting in increased movement time. This observation aligns with the intuitive understanding among squash players that better shots are designed to disadvantage opponents by prolonging the time required to reach them.

The positive relationship between movement time (MT) and ID is illustrated by the upward trend observed in the MT vs. ID graph presented in Figure 4. This graph represents the relationship for all shots except drops, lobs, and boasts. Drives are excluded from this analysis, likely due to their distinct nature as shots characterized by high velocity, which places them in a separate shot family.

The correlation coefficient between the trend line and the actual plotted values indicates a medium correlation, allowing us to infer a positive relationship between MT and ID. While the correlation is only moderate, a clear upward trend can be observed visually.

The relatively modest correlation between the trend line and the actual MT vs. ID graph may be attributed to the presence of various minor factors, as discussed earlier in the paper, that can slightly influence shot difficulty. However, these factors are too insignificant and random to be incorporated into the formulation of the ID, as evident from the clear visual trend observed in the data.
\subsection{Shot-specific relationship between movement time and index of difficulty}

The individual relationship between movement time (MT) and index of difficulty (ID) for each shot is illustrated in Figures 5, 6, 7, and 8. These figures reveal that as the ID of a particular shot increases, the movement time decreases. This phenomenon can be attributed to the fact that harder shots require the receiving player to move faster and exert more effort, enabling them to reach the ball more swiftly. However, this negative relationship does not hold true for lobs, as they are slow and defensive shots that do not exert significant pressure on the opponent. Consequently, lobs follow the expected positive relationship between MT and ID.

Moreover, the lines in the figures appear relatively straight for all shots except boasts, indicating that the index of difficulty within a particular shot type has minimal influence on movement time. This suggests that movement time remains largely consistent for each shot type, and it is only when transitioning between different types of shots that the index of difficulty starts to impact movement time. However, boasts exhibit a distinct downward slope, signifying that the index of difficulty does affect movement time for this particular shot type. This deviation from the relatively constant movement time observed in other shots can be attributed to the unique characteristics of boasts, which are two-wall shots that significantly differ from the other shot types.


\subsection{Analyzing the effort required for different shots and its practical implications}

When considering movement time within a specific shot type, there is generally minimal variation. However, the throughput, which refers to the effort required to retrieve a shot, may vary slightly based on the index of difficulty within each shot type. For instance, a drop shot with a higher index of difficulty would likely take the same amount of time to retrieve as a drop with a lower ID, but the effort (throughput) needed to retrieve the more difficult drop would be higher. Nonetheless, these variations in throughput within shot types are relatively small. The significant changes occur when transitioning between different shot types, such as drives and lobs, or drives and boasts.

Table 1 demonstrates that the throughput for picking up drives and drops was similar and considerably higher compared to the other two shots. This can be attributed to the characteristics of these shots. For drives, the player must cover a greater distance as the ball reaches the back corner, while the ball simultaneously travels at a fast pace, necessitating quick movement. Drops, on the other hand, are often targeted towards the front corner, requiring less overall movement than drives. However, drops are delicately placed shots, and their slower nature results in a shorter reaction time for the player to reach the ball. Consequently, players exert significant effort and quickly spring from their position to retrieve drops. It should be noted that the drops used in this experiment were moderately good in order to ensure the players could reach them. Thus, high-quality drops would likely require even greater throughput and may have the highest index of difficulty. This aligns with the practical understanding that drops are frequently used as attacking shots to exert pressure on opponents. Based on the finding that drives and drops require the most effort to retrieve, it can be inferred that these shots are the most effective to play. This information enables players to focus their training on effectively executing these shots and improving their movement to retrieve them.

In terms of results, lobs and boasts exhibit similar, significantly lower throughput compared to drives and drops. This is because lobs and boasts are generally defensive shots designed to provide players with an opportunity to recover from a pressured situation. As observed, lobs and boasts have relatively longer movement times compared to the other two shots, with lobs having the longest. This is due to the time it takes for these shots to reach the front wall, allowing the player ample time to retrieve them. Additionally, this provides the player who played the lob or boast sufficient time to regain a better position on the court.

\section{Conclusion}
In this work, we performed a quantitative assessment of picking up different squash shots. We aimed to investigate the most challenging shots to pick up and the shot with the longest retrieval time. The findings revealed that the drives posed the greatest difficulty, requiring significant effort to retrieve. Conversely, the lob shot had the longest retrieval time due to its slow speed and deep placement. Furthermore, during the investigation, a potentially suitable equation for measuring the index of difficulty (ID) of any shot was identified.




\bibliographystyle{splncs04}
\bibliography{ref}

\begin{thebibliography}{10}
\providecommand{\url}[1]{\texttt{#1}}
\providecommand{\urlprefix}{URL }
\providecommand{\doi}[1]{https://doi.org/#1}

\bibitem{accot1997beyond}
Accot, J., Zhai, S.: Beyond fitts' law: models for trajectory-based hci tasks.
  In: Proceedings of the ACM SIGCHI Conference on Human factors in computing
  systems. pp. 295--302 (1997)

\bibitem{accot1999performance}
Accot, J., Zhai, S.: Performance evaluation of input devices in
  trajectory-based tasks: an application of the steering law. In: Proceedings
  of the SIGCHI conference on Human Factors in Computing Systems. pp. 466--472
  (1999)

\bibitem{accot2003refining}
Accot, J., Zhai, S.: Refining fitts' law models for bivariate pointing. In:
  Proceedings of the SIGCHI conference on Human factors in computing systems.
  pp. 193--200 (2003)

\bibitem{balakrishnan2004beating}
Balakrishnan, R.: “beating” fitts’ law: virtual enhancements for pointing
  facilitation. International Journal of Human-Computer Studies
  \textbf{61}(6),  857--874 (2004)

\bibitem{cha2013extended}
Cha, Y., Myung, R.: Extended fitts' law for 3d pointing tasks using 3d target
  arrangements. International Journal of Industrial Ergonomics  \textbf{43}(4),
   350--355 (2013)

\bibitem{crossman1983feedback}
Crossman, E.R.F., Goodeve, P.: Feedback control of hand-movement and fitts'
  law. The Quarterly Journal of Experimental Psychology Section A
  \textbf{35}(2),  251--278 (1983)

\bibitem{fitts1954information}
Fitts, P.M.: The information capacity of the human motor system in controlling
  the amplitude of movement. Journal of experimental psychology
  \textbf{47}(6), ~381 (1954)

\bibitem{guiard1997fitts}
Guiard, Y.: Fitts' law in the discrete vs. cyclical paradigm. Human Movement
  Science  \textbf{16}(1),  97--131 (1997)

\bibitem{jagacinski1980fitts}
Jagacinski, R.J., Repperger, D.W., Moran, M.S., Ward, S.L., Glass, B.: Fitts'
  law and the microstructure of rapid discrete movements. Journal of
  Experimental Psychology: Human Perception and Performance  \textbf{6}(2),
  ~309 (1980)

\bibitem{jones2018review}
Jones, T.W., Williams, B.K., Kilgallen, C., Horobeanu, C., Shillabeer, B.C.,
  Murray, A., Cardinale, M.: A review of the performance requirements of
  squash. International Journal of Sports Science \& Coaching  \textbf{13}(6),
  1223--1232 (2018)

\bibitem{mackenzie1992fitts}
MacKenzie, I.S.: Fitts' law as a research and design tool in human-computer
  interaction. Human-computer interaction  \textbf{7}(1),  91--139 (1992)

\bibitem{mackenzie1993fitts}
MacKenzie, I.S.: Fitts' law as a performance model in human-computer
  interaction.  (1993)

\bibitem{mackenzie1992extending}
MacKenzie, I.S., Buxton, W.: Extending fitts' law to two-dimensional tasks. In:
  Proceedings of the SIGCHI conference on Human factors in computing systems.
  pp. 219--226 (1992)

\bibitem{mackenzie2012fittstilt}
MacKenzie, I.S., Teather, R.J.: Fittstilt: The application of fitts' law to
  tilt-based interaction. In: Proceedings of the 7th Nordic Conference on
  Human-Computer Interaction: Making Sense Through Design. pp. 568--577 (2012)

\bibitem{mcguffin2005fitts}
McGuffin, M.J., Balakrishnan, R.: Fitts' law and expanding targets:
  Experimental studies and designs for user interfaces. ACM Transactions on
  Computer-Human Interaction (TOCHI)  \textbf{12}(4),  388--422 (2005)

\bibitem{milner1992model}
Milner, T.E.: A model for the generation of movements requiring endpoint
  precision. Neuroscience  \textbf{49}(2),  487--496 (1992)

\bibitem{sleimen2013age}
Sleimen-Malkoun, R., Temprado, J.J., Berton, E.: Age-related dedifferentiation
  of cognitive and motor slowing: insight from the comparison of hick--hyman
  and fitts’ laws. Frontiers in aging neuroscience  \textbf{5}, ~62 (2013)

\bibitem{zelaznik1988role}
Zelaznik, H.N., Mone, S., McCabe, G.P., Thaman, C.: Role of temporal and
  spatial precision in determining the nature of the speed-accuracy trade-off
  in aimed-hand movements. Journal of Experimental Psychology: Human Perception
  and Performance  \textbf{14}(2), ~221 (1988)

\bibitem{zeng2012fitts}
Zeng, X., Hedge, A., Guimbretiere, F.: Fitts’ law in 3d space with
  coordinated hand movements. In: Proceedings of the Human Factors and
  Ergonomics Society Annual Meeting. vol.~56, pp. 990--994. SAGE Publications
  Sage CA: Los Angeles, CA (2012)

\end{thebibliography}
\end{document}